\documentstyle[psfig]{mn}

\newif\ifAMStwofonts

\ifoldfss
  \ifCUPmtlplainloaded \else
    \NewTextAlphabet{textbfit} {cmbxti10} {}
    \NewTextAlphabet{textbfss} {cmssbx10} {}
    \NewMathAlphabet{mathbfit} {cmbxti10} {} 
    \NewMathAlphabet{mathbfss} {cmssbx10} {} 
  \fi
  \ifAMStwofonts
    \ifCUPmtlplainloaded \else
      \NewSymbolFont{upmath} {eurm10}
      \NewSymbolFont{AMSa} {msam10}
      \NewMathSymbol{\upi}     {0}{upmath}{19}
      \NewMathSymbol{\umu}     {0}{upmath}{16}
      \NewMathSymbol{\upartial}{0}{upmath}{40}
      \NewMathSymbol{\leqslant}{3}{AMSa}{36}
      \NewMathSymbol{\geqslant}{3}{AMSa}{3E}

    \fi
  \fi
\fi 

\ifnfssone
  \newmathalphabet{\mathit}
  \addtoversion{normal}{\mathit}{cmr}{m}{it}
  \addtoversion{bold}{\mathit}{cmr}{bx}{it}
  \newmathalphabet{\mathbfit} 
  \addtoversion{normal}{\mathbfit}{cmr}{bx}{it}
  \addtoversion{bold}{\mathbfit}{cmr}{bx}{it}
  \newmathalphabet{\mathbfss} 
  \addtoversion{normal}{\mathbfss}{cmss}{bx}{n}
  \addtoversion{bold}{\mathbfss}{cmss}{bx}{n}
  \ifAMStwofonts
    \ifCUPmtlplainloaded \else
      \UseAMStwoboldmath
      \makeatletter
      \new@mathgroup\upmath@group
      \define@mathgroup\mv@normal\upmath@group{eur}{m}{n}
      \define@mathgroup\mv@bold\upmath@group{eur}{b}{n}
      \edef\UPM{\hexnumber\upmath@group}
      \new@mathgroup\amsa@group
      \define@mathgroup\mv@normal\amsa@group{msa}{m}{n}
      \define@mathgroup\mv@bold\amsa@group{msa}{m}{n}
      \edef\AMSa{\hexnumber\amsa@group}
      \makeatother
      \mathchardef\upi="0\UPM19
      \mathchardef\umu="0\UPM16
      \mathchardef\upartial="0\UPM40
      \mathchardef\leqslant="3\AMSa36
      \mathchardef\geqslant="3\AMSa3E
    \fi
  \fi
\fi 

\ifnfsstwo
  \DeclareMathAlphabet{\mathbfit}{OT1}{cmr}{bx}{it}
  \SetMathAlphabet\mathbfit{bold}{OT1}{cmr}{bx}{it}
  \DeclareMathAlphabet{\mathbfss}{OT1}{cmss}{bx}{n}
  \SetMathAlphabet\mathbfss{bold}{OT1}{cmss}{bx}{n}
  \ifAMStwofonts
    \ifCUPmtlplainloaded \else
      \DeclareSymbolFont{UPM}{U}{eur}{m}{n}
      \SetSymbolFont{UPM}{bold}{U}{eur}{b}{n}
      \DeclareSymbolFont{AMSa}{U}{msa}{m}{n}
      \DeclareMathSymbol{\upi}{0}{UPM}{"19}
      \DeclareMathSymbol{\umu}{0}{UPM}{"16}
      \DeclareMathSymbol{\upartial}{0}{UPM}{"40}
      \DeclareMathSymbol{\leqslant}{3}{AMSa}{"36}
      \DeclareMathSymbol{\geqslant}{3}{AMSa}{"3E}
    \fi
  \fi
\fi 

\ifCUPmtlplainloaded \else
  \ifAMStwofonts \else 
    \def\upi{\pi}
    \def\umu{\mu}
    \def\upartial{\partial}
  \fi
\fi

\title{Long-term Properties of Accretion Disks in X-ray Binaries: II. Stability of Radiation-Driven Warping}
\author[Clarkson, Charles Coe, Laycock]
       { W.I.~Clarkson$^{1}$, P. A.~Charles$^{1}$, M. J.~Coe$^{1}$,
       S.~Laycock$^{1,2}$ \\ 1. Department of Physics and Astronomy,
       Southampton University, SO17 1BJ, UK \\ 2. Harvard-Smithsonian
       Center for Astrophysics, Cambridge, MA 02138, USA }

\date{Accepted .
      Received ;
      in original form 
      }

\pagerange{\pageref{firstpage}--\pageref{lastpage}}
\pubyear{2002}

\begin{document}

\maketitle

\label{firstpage}

\begin{abstract}

A significant number of X-ray binaries are now known to exhibit
long-term ``superorbital'' periodicities on timescales of $\sim$ 10 -
100 days. Several physical mechanisms have been proposed that give
rise to such periodicities, in particular warping and/or precession of
the accretion disk. Recent theoretical work predicts the stability to
disk warping of X-ray binaries as a function of the mass ratio, binary
radius, viscosity and accretion efficiency, and here we examine the
constraints that can be placed on such models by current observations.

In paper I we used a dynamic power spectrum (DPS) analysis of
long-term X-ray datasets (CGRO, RXTE), focusing on the remarkable,
smooth variations in the superorbital period exhibited by SMC
X-1. Here we use a similar DPS analysis to investigate the stability
of the superorbital periodicities in the neutron star X-ray binaries
Cyg X-2, LMC X-4 and Her X-1, and thereby confront stability
predictions with observation. We find that the period and nature of
superorbital variations in these sources is consistent with the
predictions of warping theory. 

We also use a dynamic lightcurve analysis to examine the behaviour of
Her X-1 as it enters and leaves the 1999 Anomalous Low State
(ALS). This reveals a significant phase shift some 15 cycles before
the ALS, which indicates a change in the disk structure or profile
leading into the ALS.

\end{abstract}

\begin{keywords}
stars - X-rays: binaries :pulsars :accretion disks :precession periods
\end{keywords}

\section{Introduction}

In a small but growing number of bright X-ray binaries a third (or
``superorbital'') period is present in addition to the usual
periodicities. Broadly speaking, superorbital variations divide into
two observational classes, with clear, stable X-ray intensity
variations in the $\sim$30-day range forming one category (e.g. SMC
X-1, Her X-1, LMC X-4), and more quasi-periodic, longer periodicities
in the $\sim$ 50-200 day range forming the other (see Clarkson et al,
2002, hereafter paper I, and references therein). Modulations in the
longer-period class are distinctly ``quasi-periodic'' in that even
long-term monitoring datasets do not yield precise periods in their
power spectra, but a broad peak, often superposed on an apparent
``red-noise'' continuum.  Examples of these include Cyg X-2
($P\sim$70-80d; Paul, Kitamoto \& Makino 2000, hereafter PKM00),
GX354-0 ($P\sim$70d; Kong, Charles \& Kuulkers 1998), X1916-053
($P\sim$83d; Homer et al. 2001) and X1820-30 ($P\sim$171d; Chou \&
Grindlay 2001).

With the exception of X1820-30, it has become customary to interpret
superorbital periodicities from both classes as due to a precessing,
possibly warped accretion disk, with differences in amplitude due to
differences in inclination. This scenario dates from the early history
of X-ray astronomy, in which a warped disk was proposed to explain the
behaviour of Her X-1 (Gerend \& Boynton 1976, Petterson 1977). The
intense radiation pressure from the central X-ray source drives the
disk warp, and tidal forces lead to precession on the observed 35-day
cycle. The precession causes periodic obscuration of the central
source, giving rise to a modulation of the X-ray flux at the
precession period.  This model has been the subject of intense
theoretical examination in recent years (see e.g. Wijers and Pringle
1999, or for the current state-of-the-art, Ogilvie and Dubus 2001,
hereafter OD01, and references therein).

In Paper I it was argued that varying absorption could not account for
the superorbital periodicity in SMC X-1, as variation in the 1.3 -
12.1 keV band was mirrored at higher energies. Instead, the warp could
manifest itself in two ways: as a variation in uncovered emitting area
(as in Gerend and Boynton 1976), or as a variation in accretion rate
onto the neutron star.

A further manifestation of disk warping not mentioned in Paper I is
modification of disk throughput by X-ray irradiation of the
disk. X-ray irradiation from the central source influences the disk
temperature and viscosity (van Paradijs 1996, King \& Ritter 1998). In
a flat disk, self-screening strongly diminishes this effect (Dubus et
al 1999), but if the disk is at all warped, irradiation reaches beyond
the inner disk in an azimuthally asymmetric way. In this case
irradiation can significantly heat the disk (Hynes et al 2002) and is
thought to affect the rate of mass flow through the disk (Dubus et al
2003), with measurable time-varying effects on the disk temperature
profile and central source intensity.

Scenarios other than disk warping are possible, however. We merely
introduce these mechanisms here and refer the reader to section 6.2 of
Paper I for more information:- (i) Also postulated to explain the
superorbital periodicity in Her X-1 was the possibility of precession
of the magnetic axis of the neutron star (Tr\"{u}mper et al 1986; Lamb
et al 1975). Here, the variation of accretion from equator to pole
causes a torque that leads to the precession of the magnetic
axis. (ii) A third body can induce an eccentricity in the XRB orbit,
causing orbital precession and decay. The precession manifests itself
as a change in accretion rate on a period determined by the orbital
period of the XRB and the third body (Chou \& Grindlay 2001). This
process has been postulated to explain the $\sim$ 171 day period in
X1820-303 but for SMC X-1 the low amplitude of modulation of the pulse
arrival times (Wojdowski et al 1998) requires an extremely finely
tuned triple system, so this mechanism was discarded for that system
(paper I). Mechanisms for which no prototypical system have yet been
identified involve (iii) variations in the extent to which the donor
overflows its Roche Lobe as a result of stellar pulsation (see, e.g:
Wehlau et al 1992), or (iv) varying location of an accretion X-ray
bright spot (as happens in the UV in CV's, e.g: Rutten, van Paradijs
\& Tinbergen 1992).

\subsection{Accretion Disk Warping}

In Paper I we presented the techniques of examining superorbital
variations in XRB, focusing on SMC X-1, the source which shows the
clearest variation in its superorbital period. The dynamic power
spectrum (or DPS) was used to chart the behaviour of the superorbital
period. An apparently cyclic variation (on an even longer timescale of
$\sim$7 years) in the superorbital period was evident in the DPS. It
was suggested that the behaviour of SMC X-1 is consistent only with a
precessing, warped accretion disk, where a competition of warping
modes leads to variation in the warp, giving rise to the shifting
superorbital period.

This interpretation gains support from the theoretical stability
analysis of OD01, which extends $\alpha$-disk theory to
radiation-driven warping, with correct analytic expressions for the
torque on a disk element. Solutions are examined numerically to
determine stability to warping for initially flat disks, steadily
precessing disks and nonlinear warps. They find that for reasonable
values of the global disk viscosity $\alpha$ and accretion efficiency
$\eta$, the stability of an accretion disk to warping is determined by
the binary separation $r_b$ and mass ratio $q=\frac{M_2}{M_1}$. A
further variable is the location of the mass input to the disk,
whether close to the Lindblad resonance point or the circularisation
radius $r_c$. For most X-ray binaries, the majority of the torque
exerted by the mass input is close to $r_c$, as even in LMXB with
large accretion disks, the matter stream is believed to continue to
the vicinity of the circularisation radius. Accordingly, we follow
OD01 in taking the location of mass addition to be $\sim r_c$. Figure
7 shows the division of XRB into regions of stability to warping from
this analysis. The location of SMC X-1 at the boundary between regions
stable to simple mode 0 warping and combinations of modes supports our
assertion that the long-term X-ray behaviour of SMC X-1 is due to a
competition of warping modes.

The location of the boundaries between stability regions in $r_b$ -
$q$ space depends strongly on the global disk viscosity parameter,
$\alpha$ and the neutron star accretion efficiency $\eta$. Uncertainty
in these values translates into uncertainty as to the location of the
boundary between regions and thus the stability classification of a
system. However, the adopted values of $\alpha$ = 0.3 and $\eta$ = 0.1
are usually assumed to be appropriate for neutron star XRB (Frank,
King \& Raine 1995, OD01 and references therein).


In this paper we examine the long-term behaviour of a selection of
X-ray persistent neutron star XRB in which superorbital periodicities
are well-established. We apply the techniques from paper I, in
particular the DPS, to explore the phenomenology of these variations.
We focus on neutron star XRB for which persistent superorbital periods
have previously been reported. Our goal is to relate the long-term
behaviour of the sources to their position in the $r_b$ - $q$ diagram,
so as to directly confront the stability predictions of the
radiation-driven warping framework with observation. If succesful, it
will establish the time evolution of periodicities as a powerful
indicator of XRB properties in systems whose properties are less
well-understood, for example the Ultracompact X-ray Binaries (Clarkson
et al 2003, in prep)

\section[]{RXTE Observations}

Launched in December 1995, the Rossi X-Ray Timing Explorer (RXTE)
carries an All Sky Monitor (ASM), which gives regular coverage of the
entire sky.  Typically 5-10 readings - called ``dwells'' - are taken
of each of a list of sources per day, lasting about 90 seconds per
dwell. Timing information is provided to within a thousandth of a day,
as well as crude spectral information.  The ASM is sensitive to photon
energies between 1.3 and 12.1 keV, broken into three energy channels
(1.3-3.0~keV, 3.0-5.0~keV and 5.0-12.1~keV). Nearly 7 years of data
(January 1996 - August 2002) from the ASM were used in our
analysis. For all analyses reported in this paper, the ASM data were
selected by background and quality of coverage; all points with
background level above 10 $cs^{-1}$ were rejected as part of a
filtering technnique modeled after Levine et al (2000). Crude X-ray
spectral information can be obtained from the ASM data by constructing
hardness ratios of the three channels. In this work we define the
ratio to be the sum of the count rate in the two high energy channels
(3.0 - 12.1 keV) to that at low energy (1.3 - 3.0 keV).

\section{Source Selection}

Each source used in this analysis was selected on the basis of three
criteria. Firstly, independent estimates of $r_b$ and $q$ must be
available, which restricts our search to optically identified systems
with full orbital solutions. Secondly, the sources must be persistent
X-ray emitters at a high enough level to provide long-term datasets of
sufficient quality to detect long-term modulations, which favours
neutron star XRB. Third, each system used must show independent
evidence for a persistent accretion disk.

Table 1 shows the sources selected for this study, together with the
warping behaviour predicted by OD01. The strong dependence of the
stability boundaries on $\alpha$ and $\eta$ leads us to choose systems
with similar compact objects in the hope that for such systems these
parameters will take similar values. Poor coverage of the $r_b$ - $q$
diagram forces us to select systems with both high and low - mass
companions.

Figure 1 shows sections of the RXTE/ASM lightcurves from the sources
used here, binned appropriately to improve signal-to-noise. The
lightcurve of Cyg X-2 shows similar morphology to that of SMC X-1,
though at count rates about a factor of 20 higher. Its morphology is
also more complex than a single periodic cycle, which leads to the
detection of more than one periodicity (PKM00). The ASM lightcurve for
Her X-1 shows a clear variation on the well-known 35 day cycle, with a
secondary peak half a cycle after the main period of extended
activity. LMC X-4 also shows evidence of a cycle that is superficially
similar to that of SMC X-1, although the lightcurve shows more
scatter. 

\subsection{Discarded Sources}

Several well-constrained, persistent neutron star XRB are within the
regions of interest in the $r_b - q$ diagram, but we choose not to
analyse them here as they are unlikely to show significant long-term
cyclic phenomena in the RXTE/ASM. Circinus X-1 is well within the
region of the stability diagram corresponding to mode 1 and higher
warping, and is a luminous neutron star XRB (Tennant 1986). However,
its orbit is strongly suspected to be highly eccentric, among many
other properties the system shares with A0538-66 (Charles et
al. 1983). As pointed out in OD01, the stability analysis as it stands
is not valid for non-circular orbits.

4U 1907+09 is a pulsar XRB with massive companion (e.g. van Kerkwijk
et al 1989) in a wide orbit, that has shown a monotonic {\it increase}
in pulse period since its discovery (Mukerjee et al 2001). Transient
QPO's have twice been discovered at frequencies ($\sim$10 - 400 mHz),
the interpretation of which requires an accretion disk, however
small. This phenomenon suggests that temporary formation of an
accretion disk is the mechanism responsible for the spin-down, and
that this must occur during the majority of orbital cycles (Mukerjee
et al 2001). The interval over which this disk persists is small
compared to the 8.4-day orbital period (in 't Zand et al 1998),
suggesting that radiation-driven warps would not persist long enough
to be measurable by RXTE/ASM.

X2127+119 has recently been shown to consist of two sources which
cannot be resolved by RXTE (White \& Angelini 2001; Charles, Clarkson \& van Zyl 2002). While resolving the mystery of how a source can show evidence
for both an ADC and Type I bursts, it is not possible to relate
components in the observed RXTE/ASM X-ray flux to a single system.

OAO1657-415 shows spin-up and spin-down trends without any
correlation with X-ray luminosity $>$ 20keV (Baykal 2000), suggesting
no persistent accretion disk is present. In the case of OAO1657-415,
these episodes cannot be explained by wind accretion (Baykal 1997),
but are thought to arise through temporary formation and destruction
of a disk, in the manner implied by Bildsten et al (1997). A small
correlation between accretion rate and pulse frequency change in the
RXTE waveband is taken as evidence that this is indeed the case.

\small
\begin{table*}
\begin{center}

\caption{Neutron star XRB exhibiting superorbital modulations}

\begin{tabular}{lccccccl}
\hline

Name & $r_b$ & $q$ & $P_{sup}$ $^1$ & $M_X$ & $P_{orb}$ & $-\dot{M_2}$ & Prediction $^2$ \\
	& {\footnotesize ($\times 10^6 GM_1/c^2$}) & & (d) & $(M_{\odot})$ &(d) & ($\times 10^{-8}$$M_{\odot} yr^{-1}$) & \\
\hline
Her X-1 & 3.1 & 1.56 & 35 & 1.36 & 1.700 & 1.4 & Persistent mode 0 \\ 
LMC X-4 & 4.5 & 10.6 & 30 & 1.35 & 1.408 & 3.5 &Persistent mode 0 \\ 
SMC X-1 & 8.9 & 11.0 & 50 & 1.6 & 3.890 & 2.3 & Mode 0 dominant but other modes possible \\ 
Cyg X-2 & 8.0 & 0.34 & 78 & 1.78 & 9.844 & 2 & Complex combination of modes 0,1 and higher \\ 

\hline

\end{tabular}
\end{center}
{\footnotesize $^1$ Claimed superorbital period} 
{\footnotesize $^2$ Based on position in figure 7 of OD01}
{\footnotesize $^3$ See text for references}

\end{table*}
\normalsize

\begin{figure*}
\begin{center}
\psfig{file=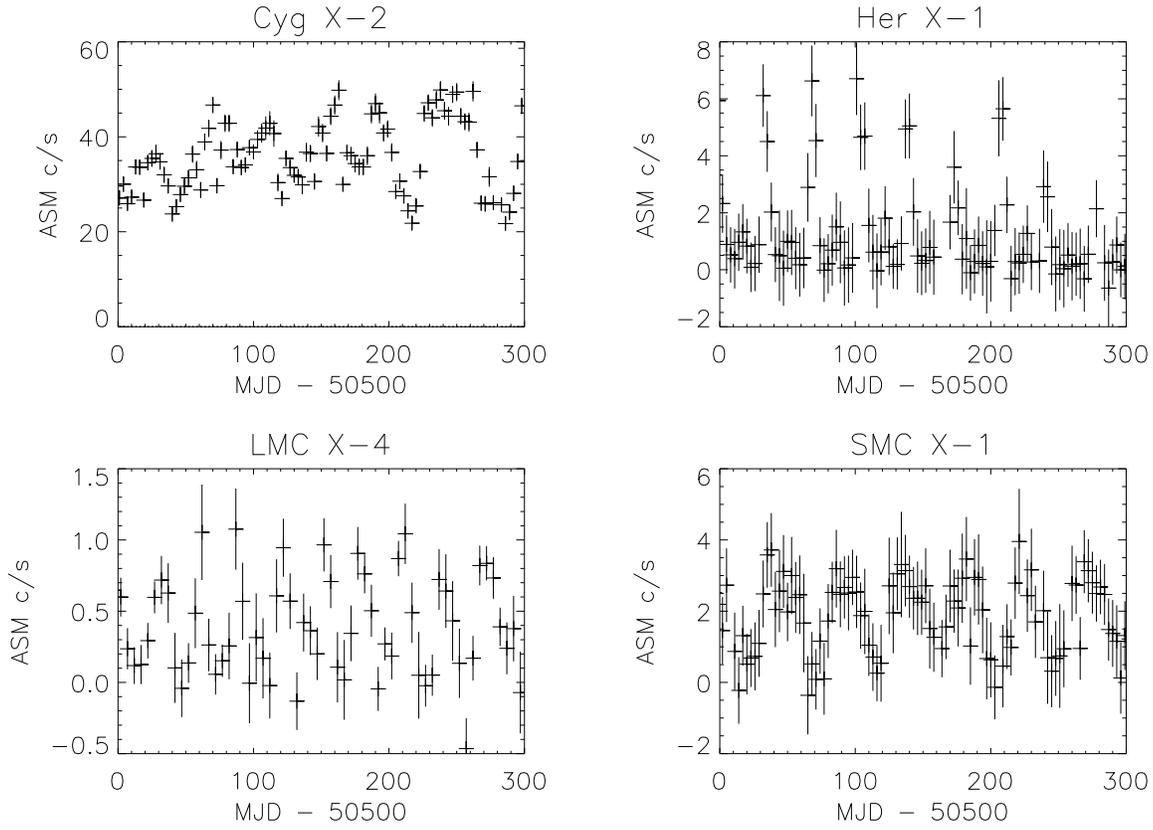,width=16cm}
\caption{Sections of the RXTE/ASM lightcurves of the sources examined in this work, binned appropriately to emphasise their long-term behaviour.}
\label{fig:bins1}
\end{center}
\end{figure*}

\section{Dynamic Power Spectrum}

Data were analysed using our DPS approach, designed
to be sensitive to long-term quasi-periodic variations. The
Lomb-Scargle (LS) periodogram code (Scargle 1982, 1989) was used in
conjunction with a sliding `data window' to produce power density
spectra (PDS) for a series of overlapping stretches of the time
series. Adjustable parameters in the analysis were the length of the
data window and the amount of time by which the window was shifted
each time to obtain the overlapping stretches of data. The code
accounted for variations in the number of datapoints per interval such
that the power spectrum resolution was identical for each interval.

Choice of data window length was influenced by the limits of reliable
period search with the PDS; we choose two complete cycles as the
minimum needed for a reliable detection. The range of claimed
superorbital periodicities runs from $\sim$ 30 - 200 days, so we set
our window length at 400 days. We choose the maximum length that meets
our criteria because this brings about the sharpest period detections.

\subsection{Statistical Significance of Detected Periods}

\begin{figure*}
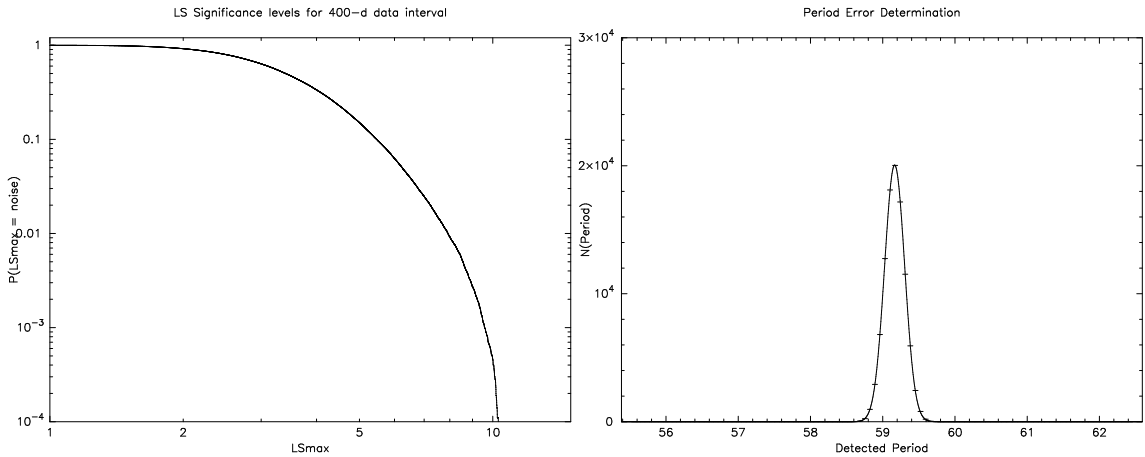

\begin{center}
\centerline{\hbox{
		\psfig{file=cygx2_cfchart_wn.ps,width=7.5cm,angle=-90}
                \psfig{file=cygx2_sigtest_pd.ps,width=7.5cm,angle=-90}
}}
\label{fig:cygx2_sigtest}
\caption{Significance and period tests for 400-day RXTE/ASM datasets. {\it Left: } Probability that a maximum power $LS_{max}$ can be generated by white-noise alone as a function of $LS_{max}$. {\it Right:} Number of trials producing detected period $P$. The period error is the $3\sigma$ width of the fit to the distribution combined with the magnitude of any shift between the most likely period detection and the input periodicity.}
\end{center}
\end{figure*}

Once a period is detected, its statistical significance must be
addressed. For most of the sources in this paper, long-term
variability is superposed on frequency-independent white-noise. The
statistical significance of a detected periodicity over this noise is
determined by computing the LS power spectra of $10^5$ simulated
400-day datasets (assuming gaussian white-noise) with the sampling of
the true lightcurve, then measuring the fraction of datasets that
produce the LS power of interest or greater (figure 2).For the 400-d
data windows used in this analysis, an LS power of 9 corresponds to a
$99.9\%$ confidence level that the detected periodicity is real, and
not a collusion of random noise and data sampling. These sources show
LS powers in excess of 30, substantially in excess of even the 99.9\%
significance level. The highly coherent nature of the variation
profiles (figure 1) argues against frequency-dependent red-noise
(e.g. Timmer \& K{\"o}nig 1995, Homer et al 2001), affording
confidence in our significance levels.

The error in the detected period is determined empirically. A
sinusoidal variation at the detected periodicity, with amplitude
corresponding to the detected peak LS power is superimposed on white
noise, with random sampling on the RXTE/ASM count rate and over the
same length of time as a data window from the DPS. The LS periodogram
of this dataset is performed and the peak frequency measured. This
process is repeated for $10^5$ simulated datasets, and the resulting
spread of detected periodicities about the input periodicity
measured. The detection accuracy is quoted in units of $3\sigma$ of
the fit to the resulting distribution in period detections. In one
case (Cyg X-2) the peak period detection was shifted from the input
period by an amount corresponding to the beat between the detected
period and $\sim$ 8 times the time interval of a data window, which
has been incorporated into the quoted error on the period detection
(table 2 and figure 2).

\section{Results}

\subsection{Her X-1}

Her X-1 is the prototypical disk warping system, with a $\sim$35-day
periodicity detected at optical and X-ray wavelengths. This
periodicity is almost certainly due to a precessing, warped accretion
disk, as first suggested by Gerend \& Boynton (1976) to explain the
optical lightcurve, and corroborated by the evolution of the pulse
profile (Scott et al 2000), and variations in X-ray illumination of
the companion HZ Her (Leahy \& Marshall 1999, Leahy, Marshall \& Scott
2000). The accretion disk is thought to be tilted and twisted (Gerend
\& Boynton 1976, Schandl \& Meyer 1994, Schandl 1996). With X-ray
pulsations at 1.24s, eclipses at 1.7 days and an A7 Roche lobe-filling
companion (Middleditch \& Nelson 1976) of $\sim$2 $M_{\odot}$, the
mass ratio and binary radius are constrained (but see Leahy \& Scott
1988 for uncertainties due to possible subsynchronous donor
rotation). On three occasions, the source has failed to reach a high
state at the time expected from the 35-day cycle (Parmar et al 1985;
Vrtilek et al 1994; Oosterbroek et al 2000). These Anomalous Low
States (ALS) last for several 35-day cycles each, and are thought to
be the result of a state change in the accretion disk (Still et al
2001).

\begin{figure}
\begin{center}
\psfig{file=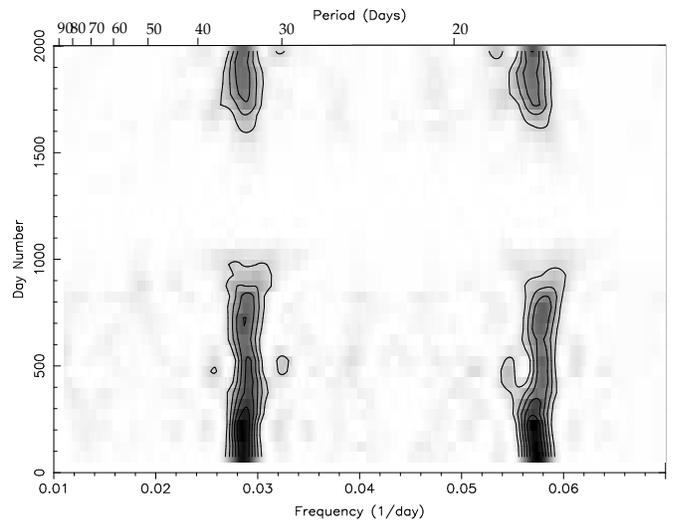,width=8.7cm}
\label{fig:dps_herx1}
\caption{DPS of Her X-1. Contours spaced at intervals of LS power 100.}

\end{center}
\end{figure}

\begin{figure}
\begin{center}
\centerline{\hbox{
                \psfig{file=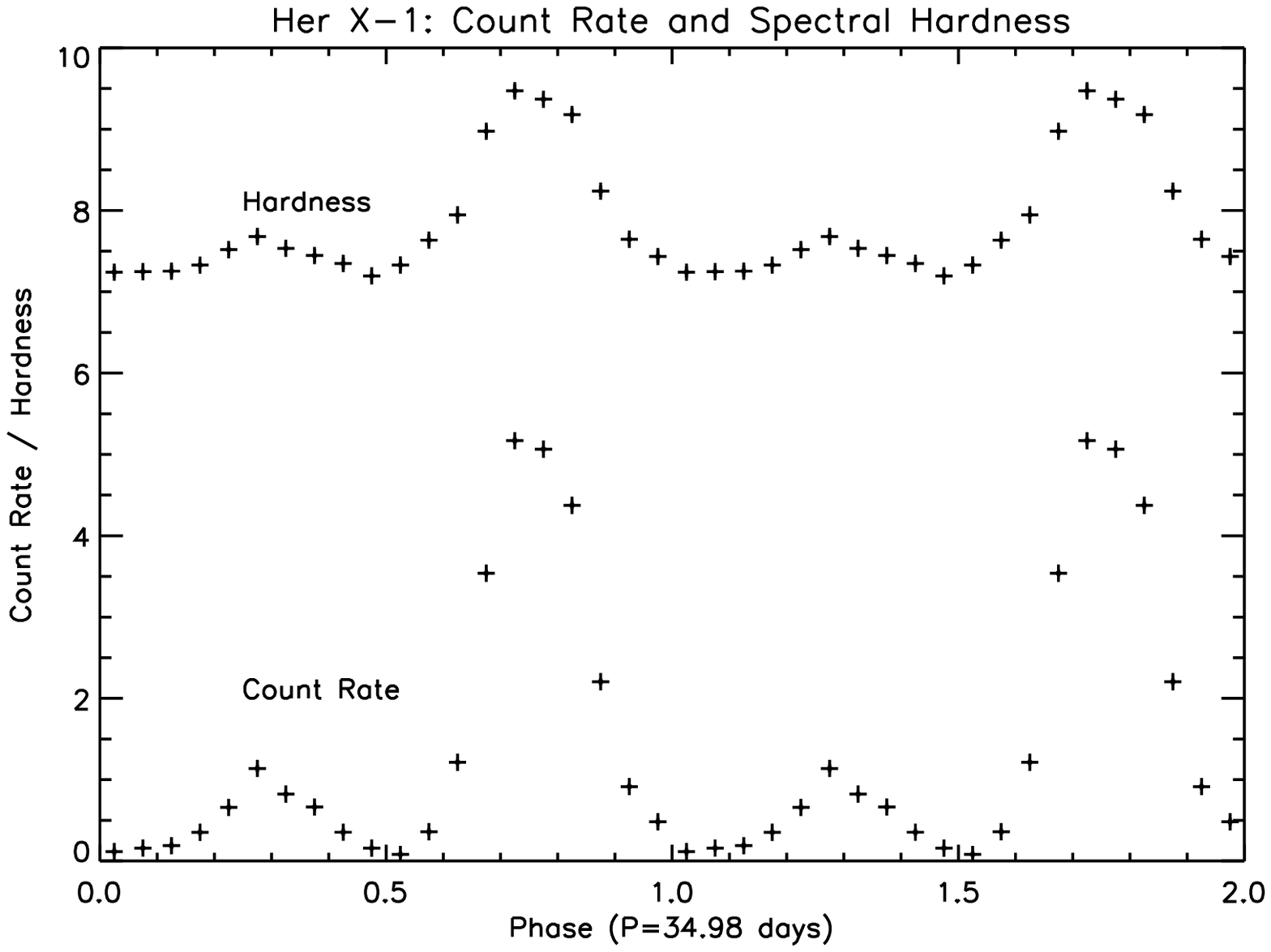,width=4.2cm}
                \psfig{file=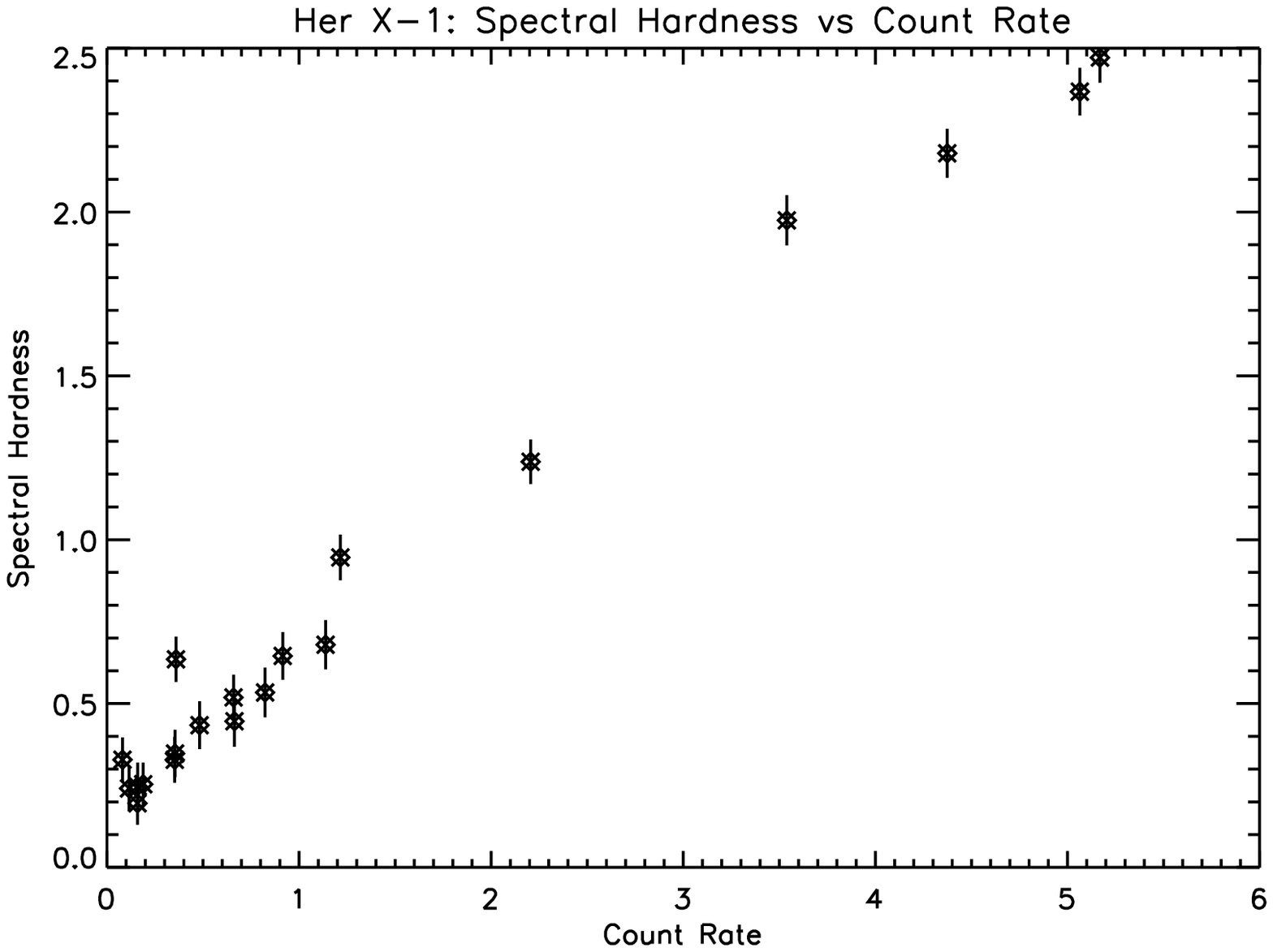,width=4.2cm} }}
\label{fig:comps_herx1}
\caption{Superorbital lightcurve ({\it left}) and count rate-hardness relation ({\it right}) for Her X-1. The spectral hardness varies with the ASM count rate and there is a clear relation between the two quantities.}
\end{center}
\end{figure}

The DPS of Her X-1 (figure 3) shows an extremely significant ($>>$99.99
\%) periodicity at 34.98 days, along with a less
significant secondary peak at 17.5 days. This secondary peak is not an
artefact of the search technique; as figure 3 shows, the DPS has
picked up the secondary peak at phase 0.3 in the 35-day cycle. Her X-1
is the only system suspected of disk precession to show this secondary
feature so significantly. Her X-1 also shows the clearest relation
between ASM count rate and spectral hardness of all the sources
studied here (figure 4).

\subsection{LMC X-4}

LMC X-4 is a $\sim$13.5 s X-ray pulsar with a 15 $M_{\odot}$ O-type
companion that eclipses the X-ray source every 1.4 days. QPO's have
been found at 2-20 mHz during large X-ray flares (Moon \& Eikenberry
2001) confirming the presence of orbiting matter at least $\sim$ 0.25
$r_{circ}$ from the neutron star. Disk modeling of the orbital and
precessional optical lightcurve (Heemskerk \& van Paradijs 1989)
suggest the accretion disk may extend out to about $4r_c$.

The DPS of LMC X-4 (figure 5) shows the simplest behaviour of the
sources analysed here: a periodicity at 30.28 days that is both stable
and persistent throughout the entire dataset. The DPS is not sensitive
to the $\sim 0.02\%$ $yr^{-1}$ systematic variation of this
periodicity reported recently by Paul \& Kitamoto (2002). At LS power
$\sim$ 38, this 30-d cycle is somewhat less significant than for Her
X-1 (LS Power $\sim$ 300), but the power still corresponds to $>
99.9\%$ significance. There appears to be slight jitter in the peak
period, but at $\pm \sim 0.05$ days this is not significant when
compared with the $3\sigma$ uncertainty for this source of $\pm 0.46$
days. LMC X-4 shows a relationship between RXTE/ASM spectral hardness
and count rate (figure 6) that is somewhat less steep than for Her
X-1.

\begin{figure}
\begin{center}
\psfig{file=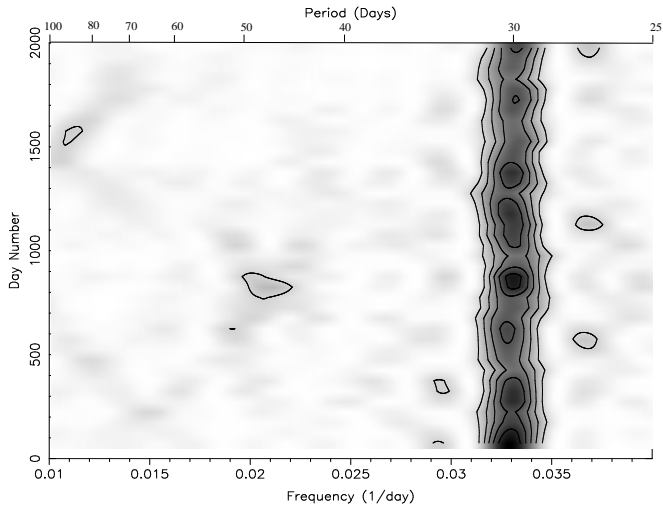,width=8.7cm}
\label{fig:dps_lmcx4}
\caption{DPS of LMC X-4. Contours spaced in intervals of LS power 10 (by comparison, LS power 9 corresponds to 99.9\% significance over noise - section 4.1)}

\end{center}
\end{figure}

\begin{figure}
\begin{center}
\centerline{\hbox{
                \psfig{file=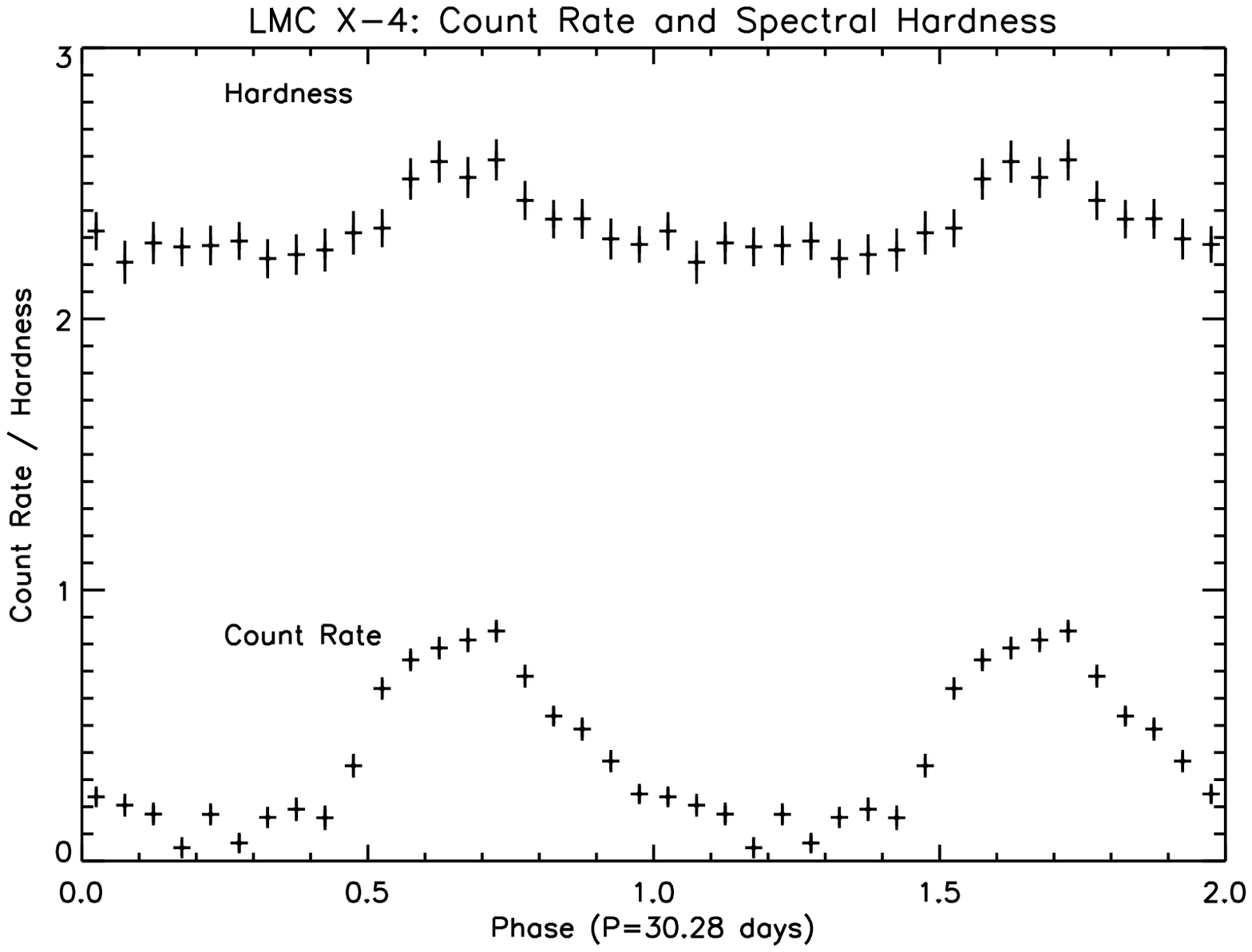,width=4.2cm}
                \psfig{file=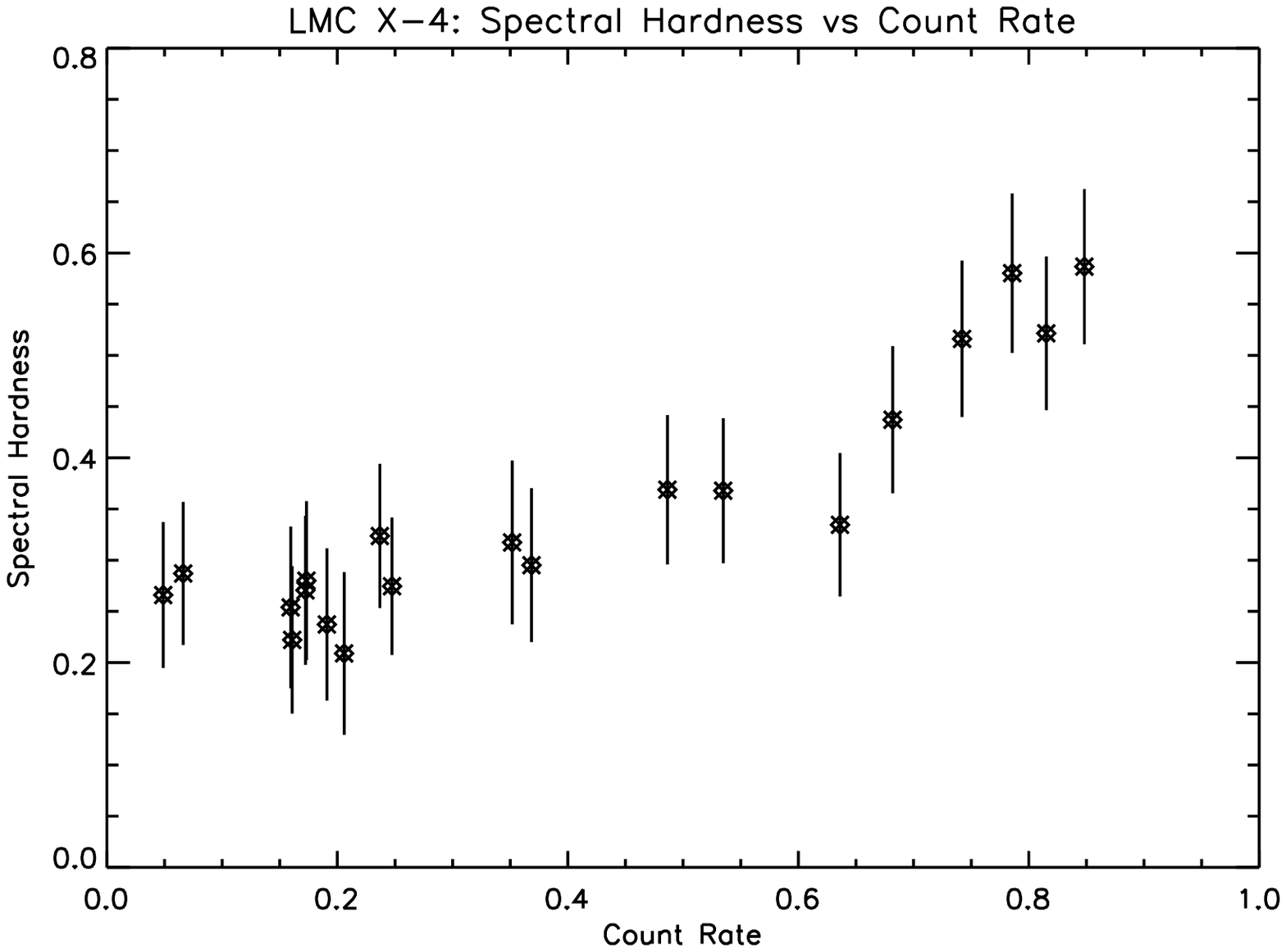,width=4.2cm} }}
\label{fig:comps_lmcx4}
\caption{Superorbital lightcurve ({\it left}) and count rate - hardness relation ({\it right}) for LMC X-4. Though a more noisy dataset, there is a correlation between the variations of the two quantities.}
\end{center}
\end{figure}

\subsection{Cygnus X-2}

Although not a pulsar, Cyg X-2 nevertheless has a well-constrained
binary orbit, with a 0.5 - 0.7 $M_{\odot}$ companion in a 9.84-day
period (Casares, Charles \& Kuulkers 1998), unusually long for a
LMXB. Type I X-ray bursts (e.g: Kahn \& Grindlay 1984) established Cyg
X-2 as a neutron star XRB. Modeling of IUE spectra (Vrtilek et al
1990) yields a best-fit outer accretion disk radius of $\sim 0.6$
$R_L$ (the Roche lobe radius), while modeling of optical lightcurves
(Orosz \& Kuulkers 1999) suggests the disk radius may be closer to 0.9
$R_L$. A pattern of high- and low-intensity X-ray states has been seen
in early RXTE/ASM, Vela 5B and Ariel-5 observations (Wijnands et al
1996), on timescales of $\sim$77.7, $\sim$77.4 and $\sim$ 77.7 days
respectively, while the Ginga/ASM lightcurve shows the strongest
periodicities on periods of 53.7 and 61.3 days (PKM00). This variation
has been attributed to radiation-driven accretion disk precession
(Vrtilek et al. 1997).

Cyg X-2 shows complex behaviour in the DPS, with an extremely
significant (LS $>$ 900) periodicity at period $\sim$ 59 days, seen
over the period from day number 600-900 (figure 7). This feature seems
to form as two separate periodicities converge, then diverge as the
feature vanishes. There are also significant recurring features at
$\sim$ 40 days. The $\sim 77$-day periodicity noted previously may be
present during the beginning of the RXTE/ASM dataset, though is of
equal or lower significance than its harmonic. In any case the
longterm variation of Cyg X-2 is not the reliable clock suggested by
Wijnands et al (1996). In contrast to Her X-1 and LMC X-4, the binned
superorbital lightcurve of Cyg X-2 shows a negative relation between
the RXTE/ASM spectral hardness and count rate (figure 8), as expected
if the superorbital lightcurve is due to an obscuration effect
(section 6.3).

\begin{figure}
\begin{center}   
        \psfig{file=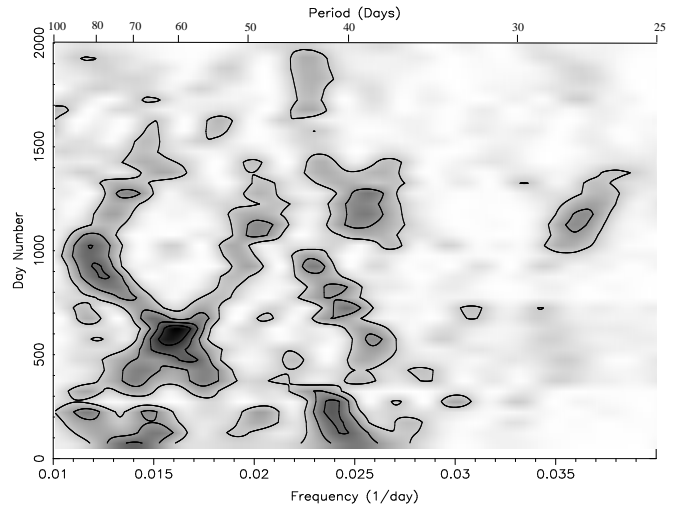,width=8.7cm}
\label{fig:dps_cx2}
\caption{DPS of Cyg X-2. Contours spaced at LS powers of 200.}
\end{center}
\end{figure}

\begin{figure}
\begin{center}  
\centerline{\hbox{
                \psfig{file=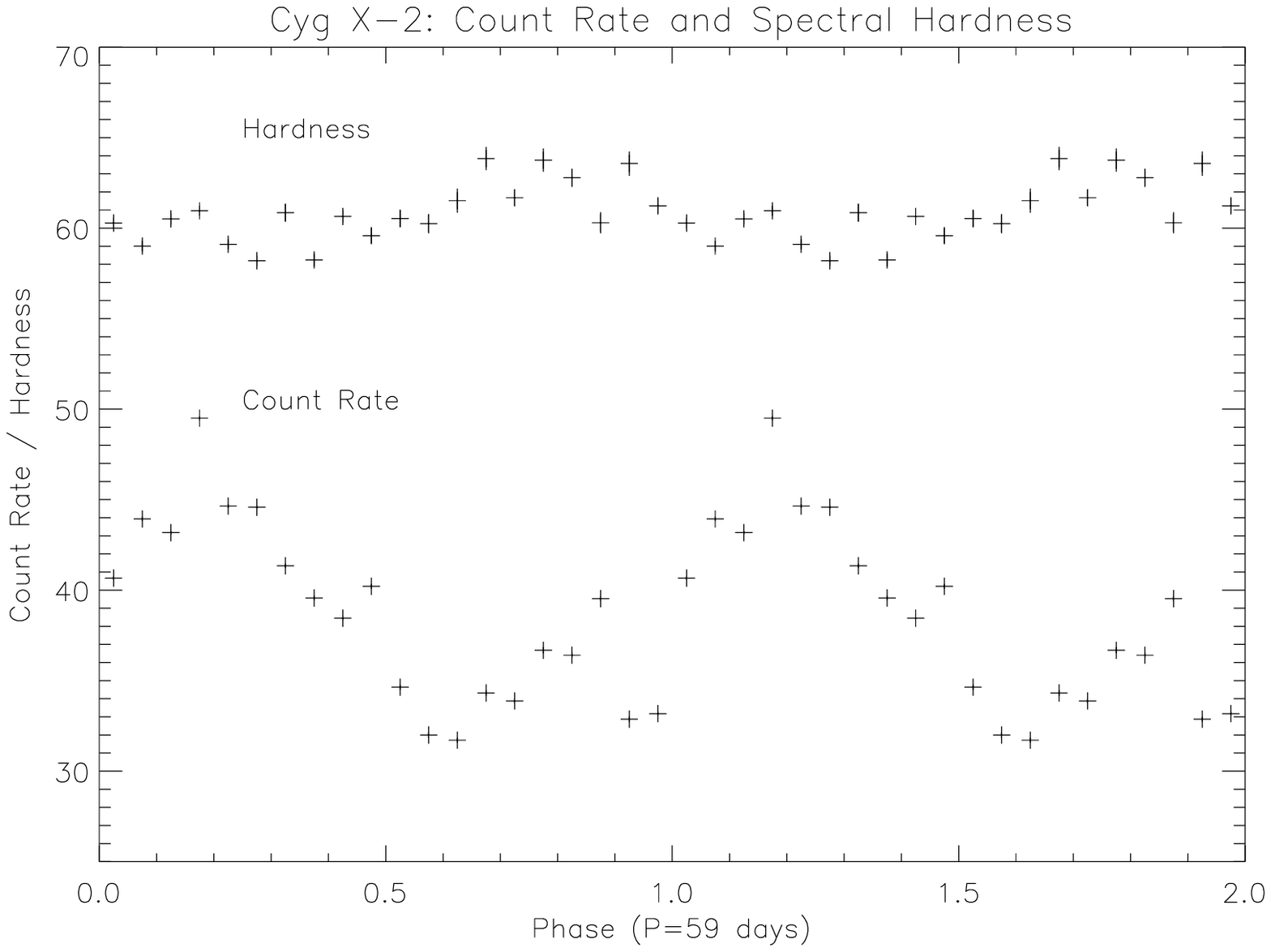,width=4.2cm}
                \psfig{file=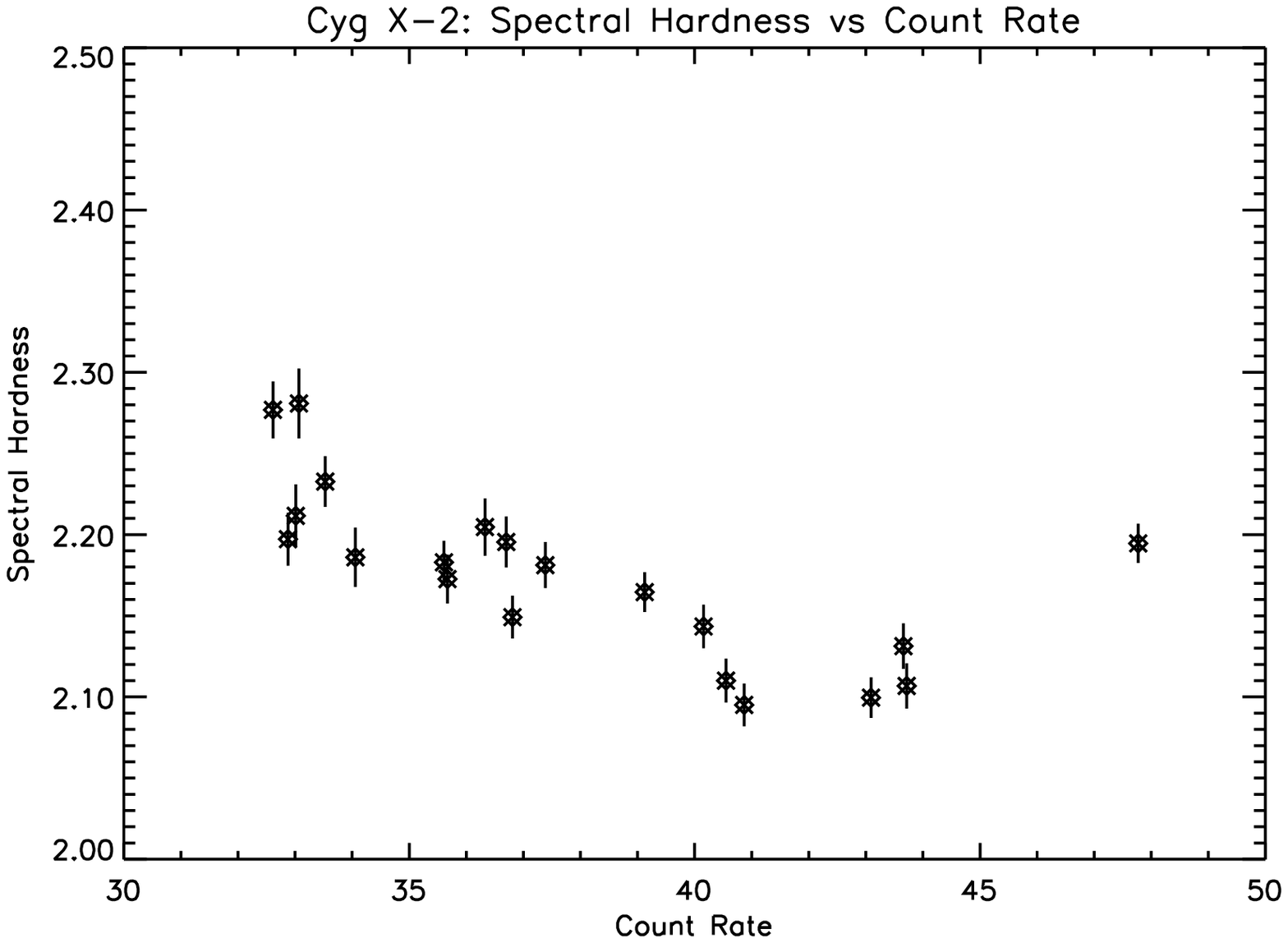,width=4.2cm} }}
\label{fig:comps_cygx2}
\caption{Superorbital lightcurve ({\it left}) and count rate - hardness relation ({\it right}) for Cyg X-2, over day numbers 600-900. In contrast to Her X-1 and LMC X-4, the spectral hardness appears to have a negative correlation with the count rate.}
\end{center} 
\end{figure}

\small
\begin{table}

\caption{Detected stable periodicities}

\begin{tabular}{lcl}
\hline

Name & $P_{sup}$ (d) & $\Delta P_{sup}$ (d)$^1$ \\
\hline
Her X-1 & 35.0 & 0.16 \\ 
LMC X-4 & 30.3 & 0.46 \\
Cyg X-2 & 59.0 & 0.45 \\ 

\hline

{\footnotesize $^1$ $3\sigma$} 

\end{tabular}
\end{table}
\normalsize

\section{Discussion}

\subsection{Her X-1: Inclination Change}
 
The clear hardness-intensity relation in the RXTE/ASM lightcurve
(figure 4) is not by itself an indication of the physical mechanism
(Tanaka 1997). In fact we find no feature of the lightcurve of this
source that contradicts the precessing, warped accretion disk scenario
of Gerend and Boynton (1976). As the warp precesses, it could uncover
regions deeper towards the central accretor, predicting the
correspondence between spectral hardness and count rate observed. The
asymmetry in the superorbital lightcurve can be brought about by a
combination of the system inclination and vertical extent of the warp
(Gerend \& Boynton 1976, Schandl \& Meyer 1994, Schandl 1996).

The superorbital periodicity is completely absent from a significant
fraction of the data, marking the third ALS detected in this source to
date (Oosterbroek et al. 2001). This is not an effect of data
sampling, as might be expected for a source with such a low count
rate; on the contrary, the persistent emission at $\sim$ 1 $cs^{-1}$
is still well sampled. Measurements by BeppoSAX (Oosterbroek et al
2001) show that the most recent ALS is accompanied by a rapid
spin-down of the neutron star, in which the spin-down rate is some
nine times faster than the spin-up rate during ``normal'' X-ray
output. At the same time, optical observations during the ALS show
(Margon et al 1999) that the companion is still strongly
irradiated. Spectral fits and lightcurve modeling of RXTE/PCA
pointings during the ALS suggest the X-ray output during the ALS
consists of X-ray reflection from the companion (Still et al
2001). The current interpretation for this behaviour is that the line
of sight to the central source has been obscured during the ALS,
causing the turn-off, by an increase in line-of-sight column density
due to the disk.

This would impact the superorbital lightcurve in two ways. Firstly,
the asymmetry between the peaks in the fold would be altered, but
whether as an increase or decrease would depend on the current
inclination and the precise disk profile. Secondly, a phase shift in
the superorbital fold would be seen during the change of inclination.
The twist of the disk brought about by the warp means that any given
feature in the warp would be seen at different disk azimuth angles for
different heights above the nominal flat disk equator.

In figure 9 we plot the {\it dynamic lightcurve} (DL) of Her X-1. In
a manner analogous to the DPS, this breaks the dataset into windows of
sufficient length to give good coverage, then folds these windows on
the ephemeris specified. Because fewer cycles per window are needed
for detection in a fold than a power spectrum, the windows can be
shorter for the DL than for the DPS; it was therefore decided to use
independent windows.

\begin{figure}
	\psfig{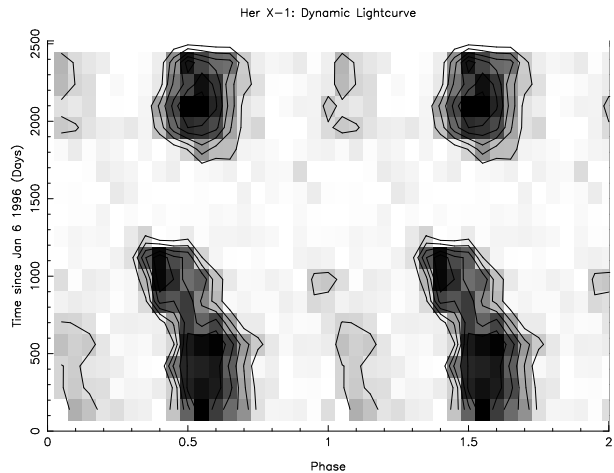}
\label{fig:newfolds_hx1}
\caption{The 35-day dynamic lightcurve of Her X-1. The phasing of the peaks undergoes a shift before the onset of the Anomalous Low State (ALS), and the relative significance of the peaks alters. When the ALS ends, the phasing is unaltered from the early years of the dataset.}
\end{figure}

From the DL we see that there is indeed a phase shift taking place
over some fifteen cycles before the ALS. The secondary peak in the
35-day cycle appears to decrease in relative amplitude just before
onset of the ALS. The cut-off in the superorbital period detection,
however, is abrupt, taking place over three cycles at most. By the end
of the ASM dataset, the system shows behaviour similar to the
beginning of the dataset, and the secondary feature is recovered.

This phase shift and profile change is most naturally interpreted as a
change in the line-of-sight disk profile. There are two good
candidates for the nature of this change. It has been suggested
(Wijers \& Pringle 1999, van Kerkwijk et al 1998) that if a disk is
sufficiently warped, the inner disk may actually flip over and start
to precess in the opposite direction to the outer disk. This would
produce the extremely rapid spin-down observed during the ALS as the
direction of flow of accreting matter would then oppose the spin of
the neutron star. Following this interpretation, we would expect to be
observing the neutron star through the accretion disk for at least
some of the ALS, which would produce a harder spectrum for the ALS
than for the normal state. However, the continuum X-ray emission is
not spectrally harder for the ALS than for the emission prior to and
after the ALS (figure 10). A further difficulty is that one would
expect to see some manifestation of the change of disk orientation in
the overall ASM lightcurve.

\begin{figure}
\psfig{file=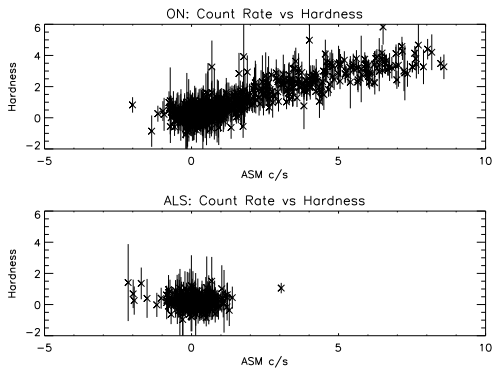,width=7cm}
\label{fig:hx1_chards}
\caption{Her X-1 Hardness-Intensity relation for the normal state ({\it top}) and the ALS ({\it bottom}), using 1-day data bins to improve S/N. There is no significant difference between the low-level emission in the two states.}
\end{figure}

A second interpretation involves the disk changing its inclination
with respect to the observer. We suppose here that the X-ray intensity
from a twisted, warped disk varies as the amount of emitting area
uncovered in the inner disk region varies. If such a disk were to
change its inclination with respect to the observer, the uncovered
area at each point in the cycle would also change with the
inclination. A twisted disk would naturally produce a variation in the
phase of features in the profile. That the ALS is accompanied by rapid
spin-down of the neutron star presents a difficulty for this scenario,
unless either (i) the accretion flow at the magnetosphere/inner disk
boundary changes direction to act against the neutron star rotation,
or (ii) the spin of the neutron star causes ejection of matter in the
manner of e.g: 4U1907+09 (section 3.1), producing a rapid spin-down as
the accretion flow lessens due to a reduction in accretion rate
through the disk. In scenario (ii), the spin-down during the ALS would
suggest that the Her X-1 disk is {\it reducing} its inclination to the
binary plane. The extent to which the uncovered area is sensitive to
the inclination change, and thus the level of other observable effects
of this change, is not at present clear. A further effect might be the
variation with the warp of the area open to X-ray heating, which might
bring about a reduction in mass throughput depending on the timescale
for such heating (c.f. Dubus et al 1999).

\subsection{LMC X-4: Stable variation}

\begin{figure}
\begin{center}
\psfig{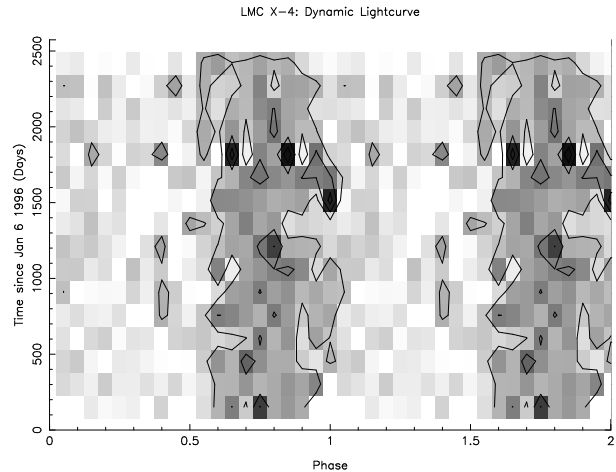}
\caption{Dynamic lightcurve of LMC X-4. Data window length chosen to include 5 cycles per window - each bin in this plot is then the average of typically 30 datapoints.}
\label{fig:lmcx4_dfold}
\end{center}
\end{figure}

At some five times more distant than most of the sources in this
study, LMC X-4 exhibits low count rate and hence lower S/N, as can be
seen in its 31-day dynamic lightcurve (figure \ref{fig:lmcx4_dfold}),
which furthermore undergoes no significant shift in phase or overall
intensity over the ASM dataset. The superorbital lightcurve itself is
somewhat different in shape to that of Her X-1 (figures 4 and 6), with
the primary peak in the profile less symmetric, and no significant
secondary peak is detected. A secondary peak might be present,
however: assuming a similar relative amplitude as that in Her X-1,
such a peak would occur at only $\sim$0.2 c/s.

This suggests that if the accretion disk in this source does shift its
inclination, as suggested for Her X-1, the LMC X-4 accretion disk
either does not do so at all on the timescale detectable with the DL,
or the disk profile and inclination are such that any existing wobble
does not impact the superorbital fold to the same extent. We can place
a rough upper limit on superorbital phase variation of features in the
dynamic lightcurve of $\sim \pm 0.1$ orbit.

\subsection{Cyg X-2: Multimode Variation}

For Cyg X-2, in which the DPS behaviour does not show a single stable
periodicity, such analysis is not possible. Indeed, its behaviour is
highly complex, with significant periodicities forming and shifting
over the 7 years of the ASM dataset. This explains the range of
modulations reported for this source over the years (e.g: Smale \&
Lochner 1992, PKM00). Of particular interest is the apparent
convergence, reinforcement and subsequent divergence of periodicities
about April 1996 - January 1999 (figure 7). Because the accretion disk
is a fluid entity and the DPS overlap between data windows of the same
order as the warp precession periods, we do not expect too
discontinuous a change in the variation shape: in other words we
expect the pattern (primary peak+single harmonic) to persist over a
timescale of several data windows.  At the point at which the strong
DPS peak appears to split into two modes, we interpret the mode
evolving towards lower frequencies as the dominant frequency, as it is
more significant. It is thus easy to interpret this feature as a
single periodicity with ill-defined phasing (hence the broad peaks in
the DPS), with a period that changes in a manner superficially similar
to SMC X-1, only on a much shorter period: $\sim$2 years compared to
the $\sim$7 years of SMC X-1.

However, the behaviour may be more complex. Closer examination of the
secondary peak at $\sim$ 40 days suggests it is not a simple harmonic
of the primary period, as its period is some 20\% longer than the
expected location of the harmonic (see also PKM00). Furthermore, this
structure is still present over the last 1500 days of the dataset,
even though the fundamental it would be a harmonic of has all but
disappeared. This feature may represent either (i) a slightly shifted
harmonic that represents a deviation from a purely sinusoidal
variation profile, or (ii) a mode of variation independent of the
$\sim$ 59-day variation.

\begin{figure}
\begin{center}
\psfig{file=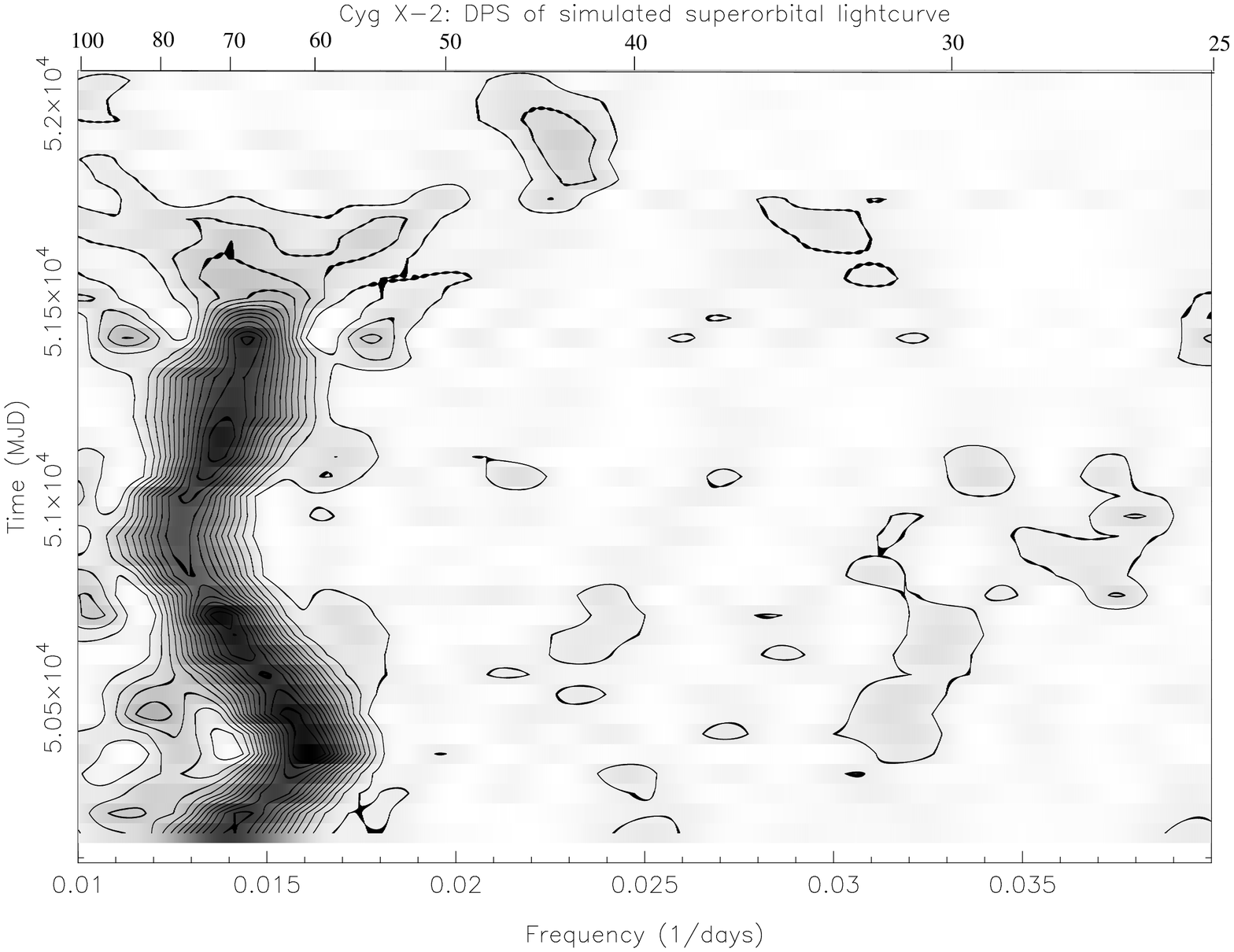,width=8cm}
\psfig{file=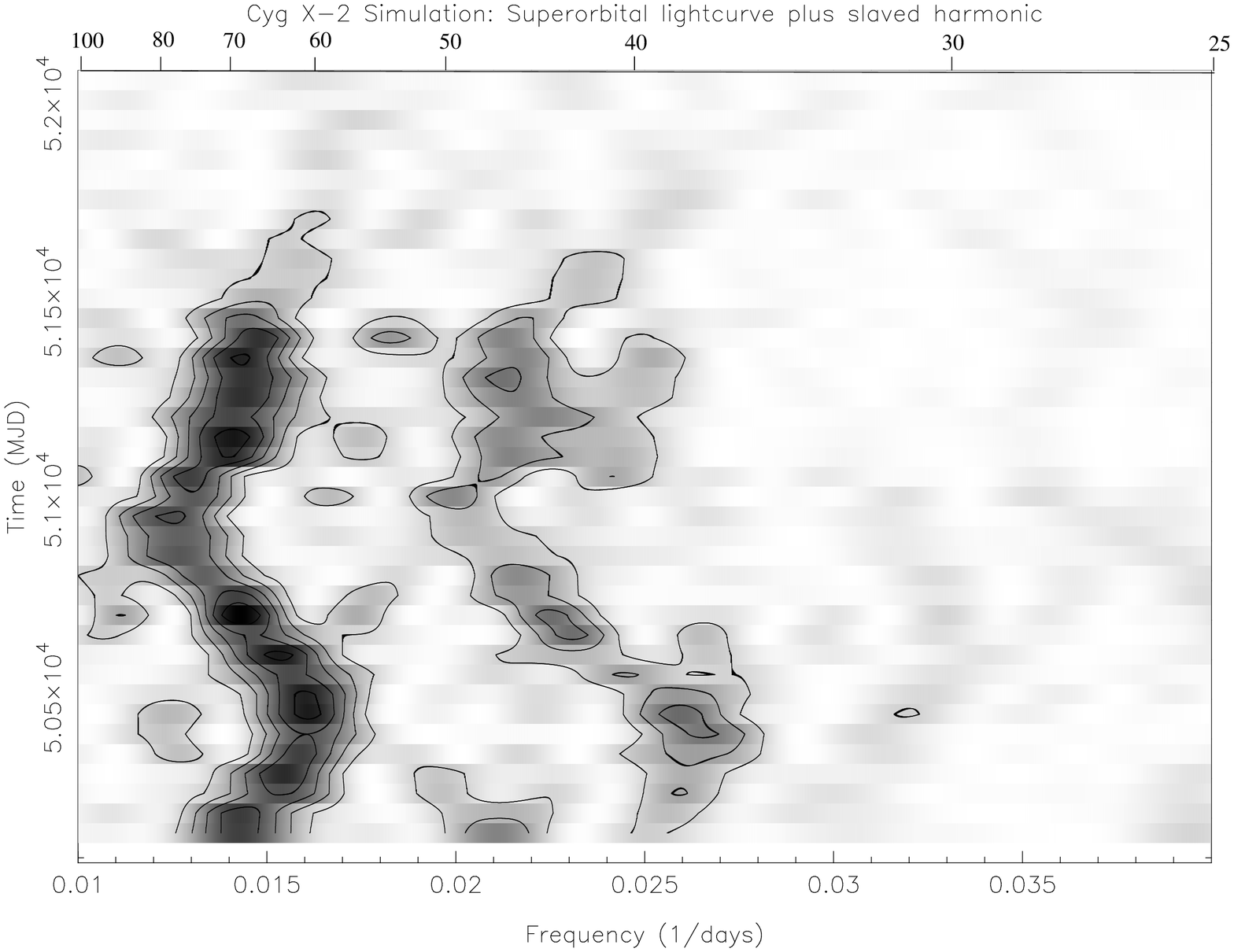,width=8cm}
\psfig{file=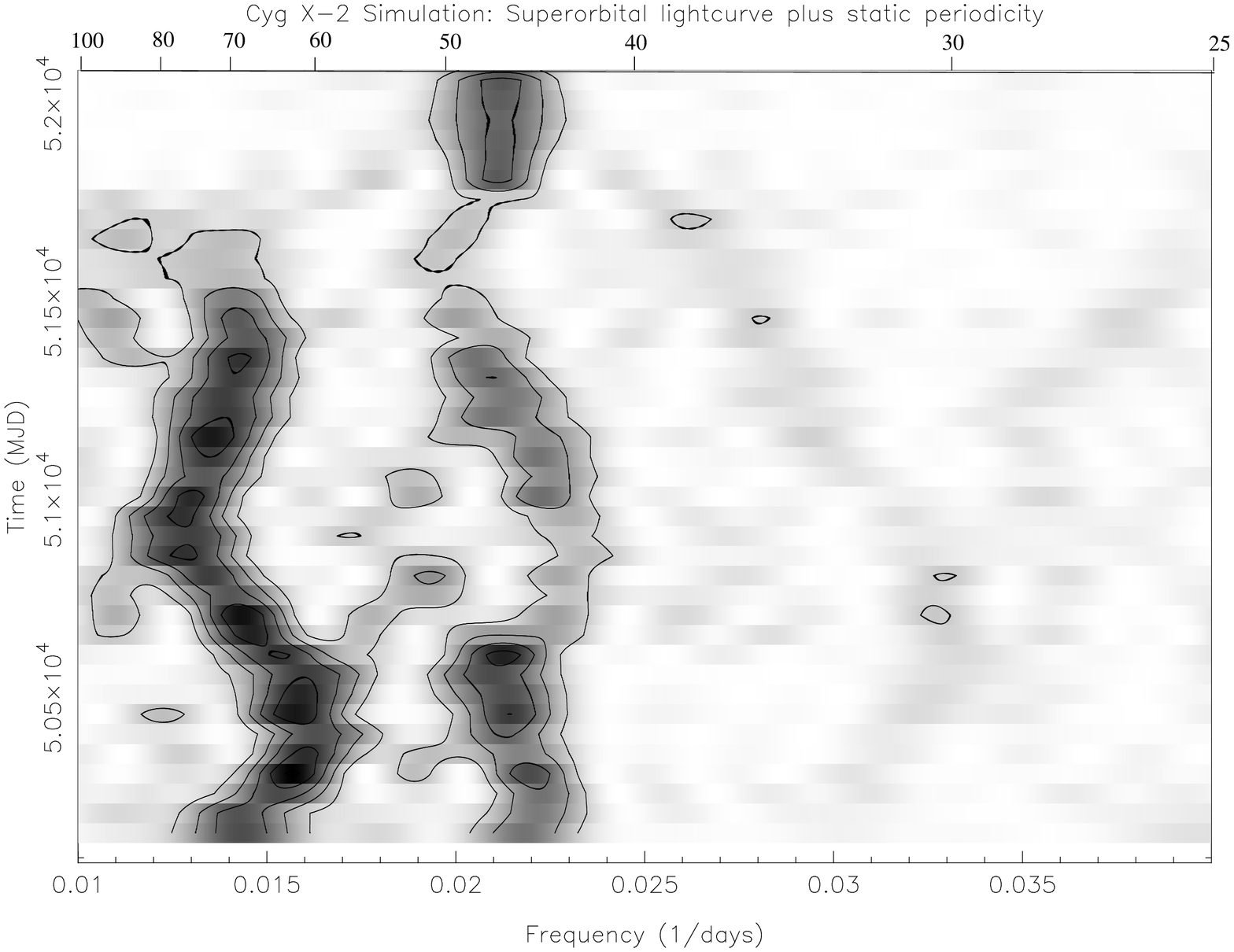,width=8cm}
\caption{DPS of simulated Cyg X-2 lightcurves. {\it Top:} input data consists of the superorbital lightcurve from day number 600-900, which undergoes period change in a manner approximating the fundamental from figure 8. {\it Middle:} As for top, but with 40-day periodicity added that is slaved to the fundamental, to approximate a non-sinusoidal shape of the variation. {\it Bottom:} 40-day periodicity kept steady as the fundamental varies.}
\label{fig:cygx2_noadd}
\end{center}
\end{figure}

We can test which scenario describes Cyg X-2 better through
simulations of varying periocicities. We take the superorbital
lightcurve over day number 600-900 (from Jan 6 1996), then scale the
period over time in a manner approximating our identified dominant
periodicity. The DPS of this simulation is quite different in
character to the real DPS (figure 12); the most obvious differences
are its far lower significance and frequency than the $\sim$ 40-day
periodicity actually seen. A 40-day periodicity was then artificially
added but slaved to the fundamental to simulate a non-sinusoidal
variation. The DPS structure that results varies more closely with the
dominant periodicity than that actually seen. If the $\sim$ 40 day
periodicity is made completely independent of the fundamental,
however, by being held steady as the fundamental varies), the DPS that
results is similar in morphology to that actually seen (figure 7).

\begin{figure}
\begin{center}
\psfig{file=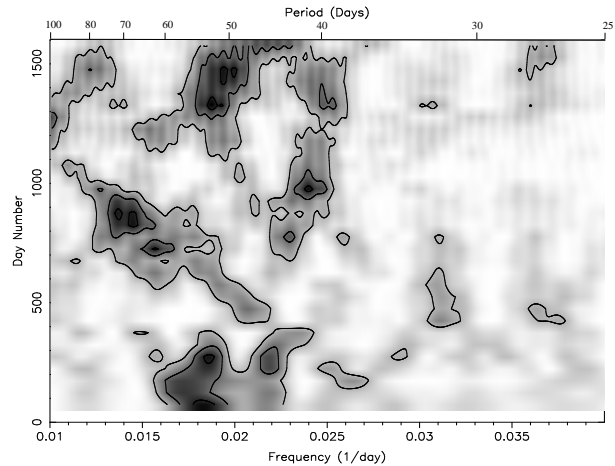,width=8cm}
\caption{Cyg X-2: DPS of the Ariel-5 dataset, binned to one-day averages, with contours at LS powers of 9 (corresponding to $99.9\%$ significance over white-noise).}
\label{fig:cygx2_ariel5}
\end{center}
\end{figure}

We thus suggest here that the apparently variable periodicity at $\sim
59$ days and the more stable periodicity at $\sim$ 40 days represent
independent variations, rather than a single variation of
non-sinusoidal shape. This suggestion is in agreement with the
conclusion of an earlier analysis based on the static periodogram of
the early years of the RXTE/ASM (PKM00). This suggestion gains some
support when we notice that the 40-day periodicity is close to the
expected harmonic of the $\sim$ 77-day periodicity reported in the
archival Vela 5B and Ariel-5 datasets. To attempt to constrain the
properties of this behaviour over a longer baseline, the Ariel-5 and
Vela-5B datasets were subjected to DPS analysis. The Vela-5B dataset
was used in the filtered, binned form provided by NASA HEASARC. Enough
points were present to use the static periodogram to confirm the
77.4-day period already noted elsewhere (e.g: Smale \& Lochner 1992),
but not enough to follow its time evolution in a statistically
significant way. However the Ariel-5 dataset has higher S/N, allowing
less coarse binning and thus providing enough datapoints to permit DPS
analysis. As figure 13 shows, the periodicities are highly variable
over time, producing the highly complex structures seen in static
periodograms (PKM00). The $\sim$ 77-day periodicity is strongly
detected over the middle $\sim$ 400 days of the dataset, while its
harmonic at $\sim$ 40 days is detected throughout most of the ASM
dataset. However, as with the RXTE/ASM dataset, there is a further
periodicity which appears variable and transient. Thus the apparent
combination of two or more separate variations does not appear to be a
new behaviour. Furthermore, this complex behaviour has now been
confirmed using two independent datasets separated by 20 years.

The X-ray manifestation of the precessing warp may be different from
that evidenced from the canonical disk precessing systems SMC X-1, Her
X-1 and LMC X-4. Cyg X-2 shows a decrease in spectral hardness with
count rate (figure 8), which at first sight would be inconsistent with
a precessing warp. However, the variation of uncovered emitting area
invoked to explain the superorbital periods in Her X-1 and LMC X-4
requires the column density through which the bulk of the X-rays reach
RXTE to remain the same as the warp precesses. If, however, the
emitting area is permanently obstructed by the outer accretion disk,
assumed not to be a significant emitter in the ASM bandpass, the
variation can be brought about by changing the column density through
which the majority of the X-rays travel. In this case, the warped disk
acts to subtract flux from the central source, rather than
contributing significantly to any emission. This removal of flux will
be seen as X-ray obscuration, which thereby causes the observed
spectrum to harden.

\subsection{Stability of Accretion Disks to Warping}

We now use our results for the other systems showing superorbital
modulations to test the stability framework of OD01. We first attempt
to identify the warping modes corresponding to the superorbital
variations identified in this work. With their stable behaviour in the
DPS and simple superorbital lightcurves, Her X-1 and LMC X-4 show
clear evidence for mode 0 warping. SMC X-1 shows a simple superorbital
fold that varies over time, probably due to disk instabilities brought
about as other modes begin to form (paper I). Cyg X-2 shows at least
two independent modes of warping, thought to represent mode 0,1 and
possibly higher warp modes.

The sources thus behave in ways qualitatively predicted by the
stability analysis of OD01 (table 1). Her X-1 and LMC X-4
are both predicted to show stable mode 0 warping, which is indeed
observed. SMC X-1 is thought to show warping commensurate with its
position near the mode 0 and mode 1 regions. Cyg X-2 is deep within
the region predicting mode 1 and higher modes, which is consistent
with the appearance and interaction of at least two warping modes.

\begin{figure*}
\begin{center}
\psfig{file=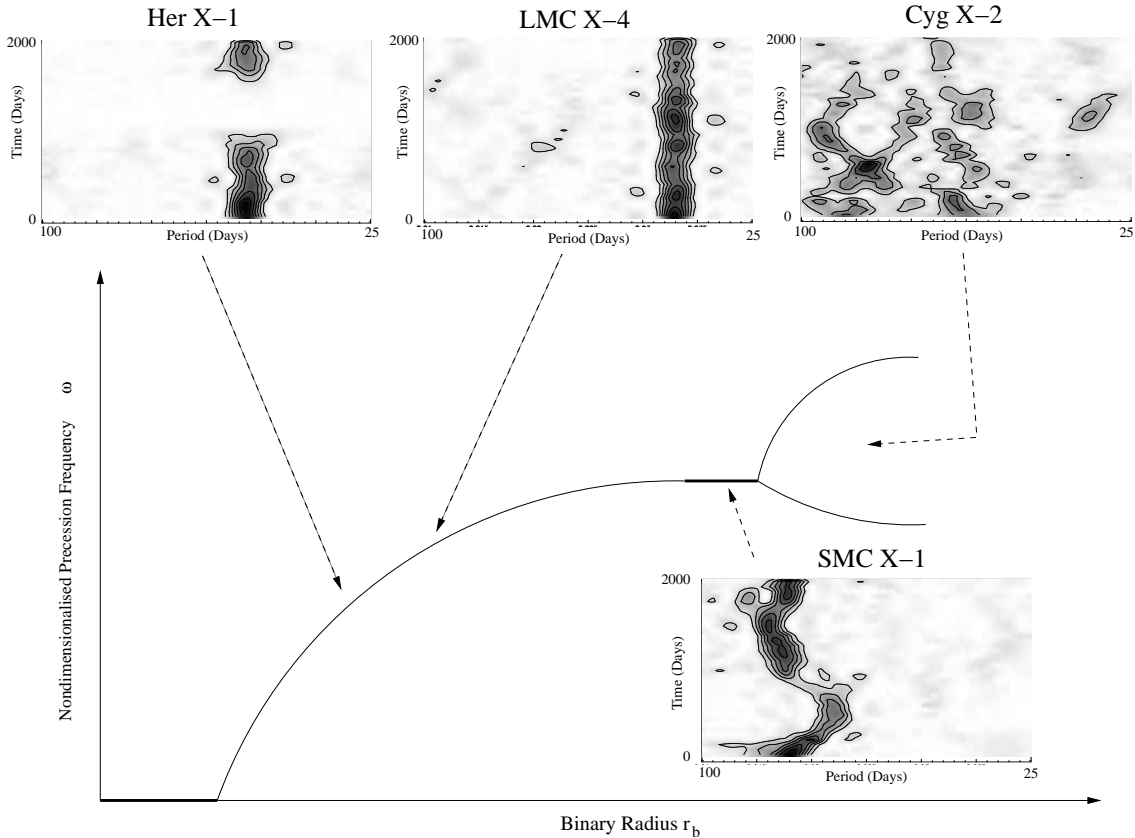,width=15cm}
\caption{Schematic bifurcation diagram for radiation-driven warping. As the control parameter $r_b$ is increased, the number of stable precession solutions increases. Initially there are none, then stable mode-0 precession rising in frequency as $r_b$ increases (solid line). Near the mode region border, the solution becomes marginally unstable as short-lived mode-1 instabilities form (thick line). Finally in the mode 1+ region, two or more steady solutions are possible, and the system precesses at a combination of these warping modes. When the positions of the sources Her X-1, LMC X-4, SMC X-1 and Cyg X-2 are overlaid on this plot, we find that their behaviour is fully commensurate with their system parameters, as shown by their behaviour in the DPS.}
\label{fig:bifurcate}
\end{center}
\end{figure*}

We illustrate the evolution of disk warping with binary radius in a
schematic bifurcation diagram, shown in figure 14. This charts the
evolution of precession solutions as the control parameter $r_b$ is
increased. According to the OD01 predictions, for low values of the
control parameter $r_b$, no {\it radiation}-driven disk warping is
predicted at present. In the region corresponding to stable mode 0
predicted warps, such warps are indeed found, with the precession
frequency increasing with $r_b$ (Her X-1 and LMC X-4 type
variations). Near the boundary between regions, modes begin to
compete, producing an instability in the superorbital period (SMC X-1
type variations). Finally, once the boundary into full multi-mode
variations is crossed, strong periodicities form and interact (Cyg X-2
type variations). When we overplot the location of these sources on
the schematic, along with their DPS results, we find that the
long-term behaviour of the sources agrees qualitatively with the
predictions of table 1, as can be determined from the DPS of the
sources.

\section{Conclusion}

We have analysed the time variation of superorbital periodicities
present in three persistent neutron star XRB: Her X-1, LMC X-4 and Cyg
X-2. Her X-1 shows a shift in phasing of the superorbital period
shortly before entering its third Anomalous Low State, which we
interpret as evidence for a change in disk inclination. LMC X-4 shows
stable mode 0 warping, with little change over the period of the
RXTE/ASM dataset. Cyg X-2 shows a combination of at least two separate
modes of variation, suggesting more complex warping shape than for the
other sources. Furthermore this behaviour has persisted for at least
$\sim$ 1000 binary orbits, as shown by new analysis of the Ariel-5
dataset.

We have identified the most likely warping modes to produce
the observed variations in Her X-1, LMC X-4 and Cyg X-2, and together
with SMC X-1 have constructed an observational stability sequence
based on their behaviour in the DPS. We find that this qualitatively
agrees with the predictions of the stability analysis of OD01 and
establishes the DPS as a powerful tool to probe XRB properties.

\bigskip

{\bf Acknowledgments}

SGTL and WIC were in receipt of PPARC research studentships. WIC
thanks Guillaume Dubus and the anonymous referee for insightful
comments. This work was only made possible through the efforts of the
ASM/RXTE team at MIT and NASA/GSFC. Archival Ariel-5 and Vela-5B data
was obtained through the High Energy Astrophysics Science Archive
Research Centre Online Service, provided by NASA/GSFC.

\label{lastpage}

\end{document}